\documentclass[sigplan,twocolumn]{acmart}
\renewcommand\footnotetextcopyrightpermission[1]{}
\usepackage[english]{babel}
\usepackage[T1]{fontenc}
\usepackage[utf8]{inputenc}
\usepackage{microtype}

\usepackage{tabularx}
\usepackage{siunitx}
\usepackage{csquotes}
\usepackage[inline]{enumitem} 
\usepackage[hypcap=true]{subcaption}

\usepackage{marvosym}   

\usepackage[normalem]{ulem}

\usepackage{listings}
\definecolor{Comments}{rgb}{0.00,0.50,0.00}
\definecolor{KeyWords}{rgb}{0.00,0.00,0.63}
\definecolor{Strings}{rgb}{0.84,0.00,0.00}

\newcommand{\mean}{\textsc{E-Mapper}}
\newcommand{\libmean}{\texttt{libMapper}}

\usepackage[
 capitalise,
 nameinlink,
 noabbrev,
] {cleveref} 

\usepackage{balance}

\usepackage{tikz}
\usetikzlibrary{calc, chains, positioning, fit, arrows.meta, decorations.pathreplacing, shapes.geometric, colorbrewer}

\pgfdeclarelayer{bg}
\pgfdeclarelayer{fg}
\pgfsetlayers{bg,main,fg}

\pgfkeys{%
  /tikz/on layer/.code={
    \pgfonlayer{#1}\begingroup
    \aftergroup\endpgfonlayer
    \aftergroup\endgroup
  },
  /tikz/.cd,
  dogear size/.initial=5pt,
}

\makeatletter

\pgfdeclareshape{dogeared}{
  \inheritsavedanchors[from=rectangle]
  \inheritanchorborder[from=rectangle]
  \inheritanchor[from=rectangle]{center}
  \inheritanchor[from=rectangle]{north}
  \inheritanchor[from=rectangle]{north west}
  \inheritanchor[from=rectangle]{north east}
  \inheritanchor[from=rectangle]{south}
  \inheritanchor[from=rectangle]{south west}
  \inheritanchor[from=rectangle]{south east}
  \inheritanchor[from=rectangle]{west}
  \inheritanchor[from=rectangle]{east}

  \savedmacro\dogearsize{%
    \edef\dogearsize{\pgfkeysvalueof{/tikz/dogear size}}%
  }

  \backgroundpath{
    \southwest \pgf@xa=\pgf@x \pgf@ya=\pgf@y
    \northeast \pgf@xb=\pgf@x \pgf@yb=\pgf@y
    \pgf@xc=\pgf@xb \advance\pgf@xc by-\dogearsize
    \pgf@yc=\pgf@yb \advance\pgf@yc by-\dogearsize
    \pgfpathmoveto{\pgfpoint{\pgf@xa}{\pgf@ya}}
    \pgfpathlineto{\pgfpoint{\pgf@xa}{\pgf@yb}}
    \pgfpathlineto{\pgfpoint{\pgf@xc}{\pgf@yb}}
    \pgfpathlineto{\pgfpoint{\pgf@xb}{\pgf@yc}}
    \pgfpathlineto{\pgfpoint{\pgf@xb}{\pgf@ya}}
    \pgfpathclose
  }
  \foregroundpath{
    \southwest \pgf@xa=\pgf@x \pgf@ya=\pgf@y
    \northeast \pgf@xb=\pgf@x \pgf@yb=\pgf@y
    \pgf@xc=\pgf@xb \advance\pgf@xc by-\dogearsize
    \pgf@yc=\pgf@yb \advance\pgf@yc by-\dogearsize
    \pgfpathmoveto{\pgfpoint{\pgf@xc}{\pgf@yb}}
    \pgfpathlineto{\pgfpoint{\pgf@xc}{\pgf@yc}}
    \pgfpathlineto{\pgfpoint{\pgf@xb}{\pgf@yc}}
    \pgfpathclose
  }
}
\makeatother

\tikzset{
    server/.style={draw, thick,  minimum width=1.5cm, minimum height=1cm, align=center},
    app/.style={draw, thick, minimum width=1.4cm, minimum height=.7cm, align=center, outer sep=0pt},
    lib/.style={densely dotted, font=\ttfamily\scriptsize\bfseries, minimum width=1.4cm, minimum height=0cm},
    dummy/.style={inner sep=0pt, outer sep=0pt},
    txt/.style={font=\tiny, align=center},
    communicate/.style={-{Stealth[round, sep=3pt, length=4pt]}},
    call/.style={-{Stealth[round, sep=3pt, length=4pt]}, densely dotted},
    active/.style={line width=3pt},
    sleeping/.style={opacity=.2, thin, on layer=bg},
    mapping/.style={draw, shape=dogeared, dogear size=4pt, minimum size=.7cm, font=\scriptsize\itshape},
    socket/.style={draw, shape=diamond, thick, minimum size=.3cm},
    read mapping/.style={-{Latex[round, sep=1pt, length=3pt]}},
    open socket/.style={-{Stealth[round, sep=0pt, length=4pt]}},
    communication/.style={{Stealth[round, sep=0pt, length=4pt]}-{Stealth[round, sep=0pt, length=4pt]}},
    message/.style={black!60, densely dotted, -{Stealth[round, sep=0pt, length=4pt]}, rounded corners=2pt}
}

\definecolor{app1bg}{HTML}{1b9e77}
\definecolor{app2bg}{HTML}{7570b3}
\definecolor{pcorebg}{HTML}{B3D7FF}
\definecolor{ecorebg}{HTML}{CCFFCC}
\definecolor{boxesbg}{HTML}{AAAAAA}

\tikzset{
    reference/.style={
      shape=circle,
      inner sep=.2ex,
      font=\footnotesize,
      text=white,
      fill=black!70,
      node distance=0pt,
    },
    listing reference/.style={
      reference,
      shape=regular polygon,
      regular polygon sides=6,
    },
}

\AtBeginDocument{
  \newcommand*{\tikzRef}[3]{\tikz[baseline=-2.6pt]{\node[reference] {#2#1#3};}}

  \newcounter{TikzCtr}[figure]
  \newcommand*{\labelTikz}[1]{\refstepcounter{TikzCtr}\theTikzCtr\label{#1}}
  \crefformat{TikzCtr}{\tikzRef{#1}{#2}{#3}}
  \Crefformat{TikzCtr}{\tikzRef{#1}{#2}{#3}}

  \newcommand*{\tikzListingRef}[3]{\tikz[baseline=-2.6pt]{\node[listing reference] {#2#1#3};}}

  \newcounter{TikzListingCtr}
  
  \crefformat{TikzListingCtr}{\tikzListingRef{#1}{#2}{#3}}
  \Crefformat{TikzListingCtr}{\tikzListingRef{#1}{#2}{#3}}
}

\usepackage{pgfplots}
\usepackage{pgfplotstable}
\usetikzlibrary{colorbrewer} 
\usepgfplotslibrary{colorbrewer}
\pgfplotsset{
    compat=1.18,
}

\begin{document}

\title{\textsc{E-Mapper}: Energy-Efficient Resource Allocation for Traditional Operating Systems on Heterogeneous Processors}
\author{Till Smejkal}
\email{till.smejkal@tu-dresden.de}
\authornote{Equal contribution from the authors.}
\affiliation{%
  \institution{TU Dresden}
  \country{Germany}
}

\author{Robert Khasanov}
\email{robert.khasanov@tu-dresden.de}
\authornotemark[1]
\affiliation{%
  \institution{TU Dresden}
  \country{Germany}
}

\author{Jeronimo Castrillon}
\email{jeronimo.castrillon@tu-dresden.de}
\affiliation{%
  \institution{TU Dresden}
  \country{Germany}
}

\author{Hermann Härtig}
\email{hermann.haertig@tu-dresden.de}
\affiliation{%
  \institution{TU Dresden}
  \country{Germany}
}

\begin{abstract}

Energy efficiency has become a key concern in modern computing. Major processor
vendors now offer heterogeneous architectures that combine powerful cores with
energy-efficient ones, such as Intel P/E systems, Apple M1 chips, and Samsung's
Exynos CPUs. However, apart from simple cost-based thread allocation strategies,
today's OS schedulers do not fully exploit these systems' potential for adaptive
energy-efficient computing. This is, in part, due to missing application-level
interfaces to pass information about task-level energy consumption and
application-level elasticity.

This paper presents \mean{}, a novel resource management approach integrated
into Linux for improved execution on heterogeneous processors. In \mean{}, we
base resource allocation decisions on high-level application descriptions that
user can attach to programs or that the system can learn automatically at
runtime. Our approach supports various programming models including OpenMP,
Intel TBB, and TensorFlow. Crucially, \mean{} leverages this information to
extend beyond existing thread-to-core allocation strategies by actively managing
application configurations through a novel uniform application-resource manager
interface. By doing so, \mean{} achieves substantial enhancements in both
performance and energy efficiency, particularly in multi-application scenarios.
On an Intel Raptor Lake and an Arm big.LITTLE system, \mean{} reduces the
application execution on average by \qty{20}{\%} with an average reduction in
energy consumption of \qty{34}{\%}. We argue that our solution marks a crucial
step toward creating a generic approach for sustainable and efficient computing
across different processor architectures.

\end{abstract}

\date{}
\maketitle

\pagestyle{plain}
\section{Introduction}
\label{sec:introduction}

With the introduction of Intel Alder Lake processors~\cite{intel_alderlake},
Intel follows the trend of heterogeneous CPUs, similar
to the processor design seen in Arm big.LITTLE~\cite{arm_biglittle} and
Apple M1~\cite{apple_m1} for the domain of powerful x86 desktop computers and
servers. Recently, AMD revealed that upcoming processor releases will also
include heterogeneous versions~\cite{amd_pheonix_2,amd_heterogeneous_cpus}. With
this, all major processor vendors now include heterogeneous CPUs in their
portfolio.

Heterogeneous CPUs typically combine a small number of high-performance cores
with a larger number of energy-efficient ones on one System-on-Chip (SoC).
High-per\-for\-mance cores offer higher single-thread performance at the cost of
increased power consumption, while energy-efficient cores have lower
single-thread performance but substantially reduce power consumption. This
difference in execution characteristics requires revisiting strategies for
resource management in modern operating systems.

The type of core to which an application thread is scheduled can significantly
affect its performance. Beyond differences in single-thread performance among
core types, some cores may support different instruction sets. For instance, the
energy-efficient E-cores in Intel Alder Lake processors lack the \emph{AVX-512}
extensions available in the  P-cores~\cite{alderlake_missing_avx512}. However,
to what extent an application can profit from high-performance cores depends on
its characteristics. For memory-bound and I/O-bound applications, for instance,
the performance gap between core types can be negligible.

User-level requirements on application importance must also be accounted for in
resource allocation. For instance, a foreground application like a video player
is typically more important than a background update. The OS should allocate
high-performance cores to the video player, even if other applications might
benefit from these cores. Moreover, application priorities can change
dynamically; for instance, an update process becomes more critical if explicitly
triggered by the user. Ideally, this priority change should be immediately
signaled to the OS, allowing prompt core reassignment and faster response to
user requests.

Modern OS schedulers have begun addressing the challenges posed by heterogeneous
cores. Windows~11 added support for the Intel Thread Director to manage process
placement on different CPU core types~\cite{windows11_scheduler,
intel_td_white_paper}. Similarly, Linux introduced the Energy-Aware-Scheduler
(EAS) extension in version~5.0, which considers the individual core performance
when assigning them to processes in  Arm's big.LITTLE systems~\cite{linux_eas}.
In addition, the Linux community is working to enhance the default system
scheduler with feedback from the Intel Thread Director~\cite{linux_td,
linux_classes}. Intel Thread Director allows for better runtime decisions on
where to allocate applications, as live characteristics of the application
behavior are taken into account. However, these systems lack an API for
applications to influence placement decisions beyond explicit core pinning.
State-of-the-art strategies used in traditional operating systems rely on simple
and fast heuristics, often resulting in unpredictable and suboptimal execution.

Optimizing the mapping of applications onto heterogeneous multi-core systems is
a well-known problem in the embedded domain~\cite{singh13survey}. Solutions can
be classified based on when the decision is taken, e.g., at compile-time,  at
runtime, or a combination of both (\emph{hybrid}). Compile-time methods offer
near-optimal mappings but lack adaptability to changing workloads, making them
unsuitable for dynamic systems where workload changes are unpredictable. Runtime
methods, in turn, consider the current system workload and generate mappings
\emph{on-the-fly} --- at application launch or during its execution. Hybrid
application mapping (\emph{HAM}) approaches combine the benefits of both
design-time and runtime methods~\cite{pourmohseni20}. HAM methods offload
compute-intensive calculations to compile-time, generating a set of
Pareto-optimal mappings for each application, and adapt these intermediate
solutions to the current workload at runtime. Modern HAM approaches consider
heterogeneous processors to optimize energy efficiency while meeting certain
Quality of Service requirements~\cite{khasanov_date20,khasanov21,
weichslgartner18,wildermann15,spieck22}.

Our work introduces \mean{}, a runtime system designed to efficiently manage
applications on desktop systems, servers, and mobile devices equipped with
heterogeneous processors. Drawing from the embedded systems domain, our approach
generalizes from HAM-based approaches, expanding them to support the execution
of both known and unknown applications -- a scenario not commonly addressed in
embedded systems. Moreover, \mean{} accommodates user-defined priorities and
supports dynamic changes in core assignments at runtime in response to
application triggers. In this paper, we also discuss the modifications that are
necessary to effectively support applications in \mean{}. Specifically, we
detail how we adapted application models such as OpenMP, Intel Thread Building
Blocks, and TensorFlow for efficient utilization in an OS that integrates
\mean{}. Furthermore, we explain how \mean{} can automatically learn and predict
the application behavior at different configurations -- a feature that is
crucial to efficiently manage a wide variety of applications.

To our knowledge, \mean{} is the first system capable of semi-automatically
managing applications with dynamic properties on heterogeneous processors while
accounting for user requirements and system triggers. \mean{} aims to fill
important gaps in modern operating systems, offering better system utilization,
improved system performance, reduced energy consumption, and enhanced user
experience. Our evaluation shows that \mean{} could reduce the energy
consumption by \qty{34}{\%} and the execution time by \qty{20}{\%} over all
applications measured on the Arm Odroid XU3-E board and the Intel Raptor Lake
Core i9-13900K.

The remainder of the paper is structured as follows: \cref{sec:background}
provides background on heterogeneous processors, their benefits, and associated
challenges. \cref{sec:design} presents the design of \mean{}, followed by
\cref{sec:selection_algorithm} outlining the resource allocation approach and
\cref{sec:runtime_dse} detailing the runtime refinement of application
configurations. \Cref{sec:evaluation} presents the evaluation of \mean{}. In
\cref{sec:related_work} we discuss other solutions to the scheduling problem on
heterogeneous CPUs, and \cref{sec:conclusion} concludes our work.

\section{A Case for Heterogeneity}
\label{sec:background}

Heterogeneous computing has become mainstream in the last years. Various types
of heterogeneous systems, including  CPU-GPU, CPU-FPGA, CPU-SmartNIC, and
CPU-ASIC combinations, have their individual strengths and weaknesses as well as
their own scaling and management challenges. In this work, however, we focus on
heterogeneous processors where different core types, all running the same
Instruction Set Architecture (ISA), are combined on one socket or System-on-Chip
(SoC).

\subsection{Heterogeneity in one Processor}
\label{sec:background:heterogeneity_in_one_processor}

Heterogeneous processors have long been used in mobile and embedded devices. The
Arm big.LITTLE architecture was the first commercial design for such processors,
consisting of two tightly connected islands of processors -- one island contains
one or many high-performance cores, and the other contains one or many
energy-efficiency cores~\cite{big.little}. The high-performance (big) cores
feature high single-thread performance and operate at a higher power envelope.
The energy-efficient (LITTLE) cores, on the other hand, feature much lower
single-thread performance at a lower power envelope. The big.LITTLE architecture
was later extended by Arm in the DynamIQ design, allowing more flexible
configurations and more core types~\cite{dynamiq}. For instance, the recent
DynamIQ Shared Unit-120~\cite{dsu120} allows combining up to three different
core types in a single processor.

In 2020, Apple released their M1 chip, which is based on Arm and uses the
big.LITTLE approach with two types of cores on one socket. The Apple M1 chip
features one island with P-cores, equivalent to Arm's big cores, and another
island with E-cores, equivalent to Arm's LITTLE cores. Later versions of the M1
chip, as well as the M2 and M3 chips, continue this design with varying
combinations of P and E-cores.

A similar design is found in modern Intel x86 CPUs, such as the Alder Lake CPUs
released in 2021 and the latest Raptor Lake CPUs released in 2022. In Core i5
versions and higher, these CPUs also feature a two-island design with P-core and
E-core islands, following an approach similar to the Arm big.LITTLE
architecture.

\subsection{Benefits of Heterogeneous Processors}
\label{sec:background:heterogenous_processors_benefit}

In the following, we discuss what trade-offs are opened up by heterogeneous
processors.

\subsubsection*{Trading Performance against Energy}
\label{sec:background:perf_vs_energy}

Modern processors, especially high-performance CPUs, have high static power
consumption due to significant leakage resulting in less energy proportionality
when scaling frequencies and voltages. Consequently, applications that cannot
leverage the full single-thread performance of individual cores waste
considerable energy on high-performance cores. This is typical for memory and
I/O-bound applications. For such applications, energy-efficient cores with less
static power are more suitable. By combining powerful and energy-efficiency
cores in one socket, operating systems can decide at runtime where to run
applications.

\subsubsection*{Trading Performance against Space}
\label{sec:background:perf_vs_space}

High-performance P-cores on modern Intel Raptor Lake processors require
significantly more space on the die than the energy-efficient E-cores, even
though the number of E-cores is larger than that of P-cores. This creates
interesting trade-offs for chip designers aiming to produce processors tailored
for certain workloads, such as CPUs with many E-cores for I/O-heavy
applications. However, these tailored systems can only be truly leveraged by
operating systems that are aware of application characteristics and can balance
the allocation of P-cores and E-cores effectively.

\subsection{Challenges of Heterogeneous Processors}
\label{sec:background:heterogeneous_processors_difficult}

\begin{figure}
    \centering
    \begin{subcaptionblock}{.48\columnwidth}
        \centering
        \includegraphics[width=\textwidth]{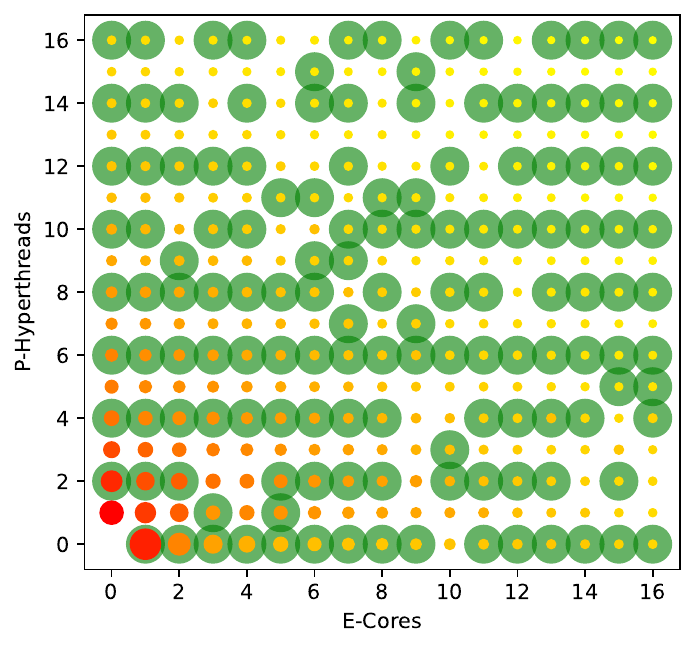}
        \caption{ep.C}
        \label{fig:background:behavior_epc}
    \end{subcaptionblock}
    \hfill
    \begin{subcaptionblock}{.48\columnwidth}
        \centering
        \includegraphics[width=\textwidth]{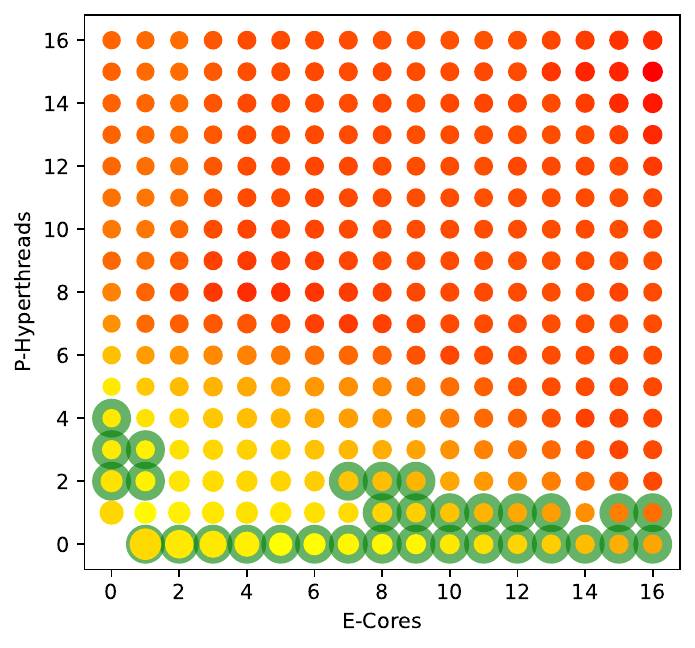}
        \caption{mg.C}
        \label{fig:background:behavior_mgc}
    \end{subcaptionblock}
    \caption{
    Performance and energy consumption of two applications on an Intel Raptor
    Lake Core i9-13900K with varying configurations. The x-axis refers to
    E-cores used, the y-axis to P-Hyperthreads. Each dot represents an
    application configuration with its thread distribution across P and E-cores.
    The dot size reflects the total execution time, and the color indicates
    energy consumption (red for higher consumption). Configurations highlighted
    in green are utilized by \mean{}.
    }
    \label{fig:background:behavior_applications}
\end{figure}

\begin{figure*}[htb]
    \centering
    \input{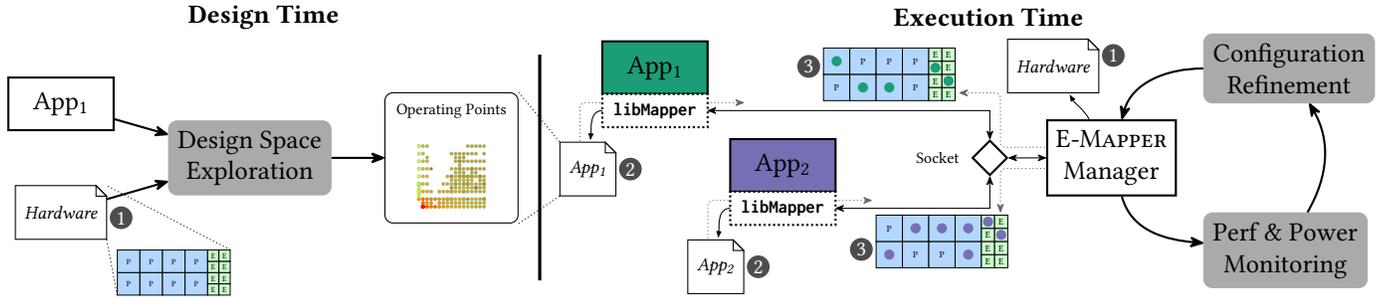}
    \caption{
      Overview of the \mean{} application management approach and system design.
    }
    \label{fig:flow}
\end{figure*}

To illustrate the performance-energy trade-off and how it varies from
application to application consider the plot in
\cref{fig:background:behavior_applications}. The figure shows different
configurations for two applications from the NAS Parallel Benchmark suite on an
Intel Raptor Lake Core i9-13900K. For \texttt{ep.C}
(\cref{fig:background:behavior_epc}), there is a smooth gradient towards the
upper right corner, both in terms of execution time and energy, indicating it
runs well on both core types and scales with increasing core counts. Conversely,
\texttt{mg.C} (\cref{fig:background:behavior_mgc}) shows degraded performance
and energy consumption on a heterogeneous core combination. Instead, this
application performs much better on homogeneous configurations with a low
overall core count.

Traditional system schedulers do not account for these differences. Moreover,
there is no interface for applications to \emph{inform} the operating system
about their workload characteristics. \mean{} addresses this by configuring
applications to use only (near-)optimal configurations, highlighted in green in
\cref{fig:background:behavior_applications}. For instance, \mean{} will never
execute \texttt{mg.C} in the less efficient red configurations in the upper
right corner, even if executing alone. On the other hand, \texttt{ep.C} can use
the upper right configuration when controlled by \mean{}, as this configuration
has benefits under certain circumstances. All this information is communicated
within the \mean{} system using a well-defined interface.

\section{\mean{} Design}
\label{sec:design}

This section details the \mean{} system's architecture and the various
application types it supports. Figure~\ref{fig:flow} provides an overview of our
approach, divided into design time and execution time components. At design
time, a high-level application description file containing the application
\emph{operating points} is created. Each operating point represents a
combination of application configuration, specific hardware resource allocation,
and non-functional characteristics such as expected utility (e.g.\ instructions
per second or transactions per second) and average power
consumption~(\cref{fig:system_overview:app_map}). This characterization of
operating points can be done through various methods, from sophisticated
design-space exploration using models, traces, and static
analysis~\cite{mariani12,haehnel_euf,castrillon2012} to simple measurement-based
resource annotations.

The core part of \mean{} lies in its execution time management approach. During
this phase, \mean{} selects one of the available operating points for each
managed application from the provided application description
files~(\cref{fig:system_overview:app_map}), optimizing a user-selected global
optimization target. For each application, \mean{} allocates corresponding
hardware resources of the heterogeneous processor and reconfigures the
applications accordingly~(\cref{fig:system_overview:mapping} + \tikzRef{}{}{4}).
The resource allocation process considers runtime demands of applications and
users, optimizing resource usage, application performance, and overall system
energy consumption.

\subsection{The \mean{} System Architecture}
\label{sec:system_architecture}

\mean{} includes three main components to interoperate with the operating system
and client applications:
\begin{enumerate*}[label=(\arabic*)]
    \item the \mean{} resource manager,
    \item a shared library (\libmean{}) for application-level integration, and
    \item application descriptions files.
\end{enumerate*}
The system design is outlined on the right of \cref{fig:flow}.

There is a single instance of the central resource manager and as many instances
of the \libmean{} library as managed applications are in the system. The \mean{}
resource manager maintains an overview of the managed applications and their
allocated resources. The \mean{} resource manager does not replace a traditional
OS scheduler but manages resource allocations and application configurations. It
searches and selects new application configurations whenever a new application
starts in the system, a running application exits, or system-level triggers
occur. The selection process, guided by the
application~(\cref{fig:system_overview:app_map}) and
hardware~(\cref{fig:system_overview:hardware_map}) descriptions, optimizes the
resource allocation for individual applications. \Cref{sec:selection_algorithm}
discusses the resource allocation optimization approach. The \mean{} resource
manager also monitors performance characteristics and power usage of all
applications to train and calibrate the application mapping descriptions at
runtime, detailed in \cref{sec:runtime_dse}.

The \libmean{} library establishes communication between the global resource
manager and individual applications and applies the configuration selected by
the \mean{} manager. Our system uses
\texttt{protobuf}\footnote{\url{https://github.com/protocolbuffers/protobuf}}
messages over Unix sockets for a flexible and extensible communication
interface. How the configuration is applied in a specific application depends
heavily on the application type and its available level of elasticity, discussed
in \cref{sec:client_types}.

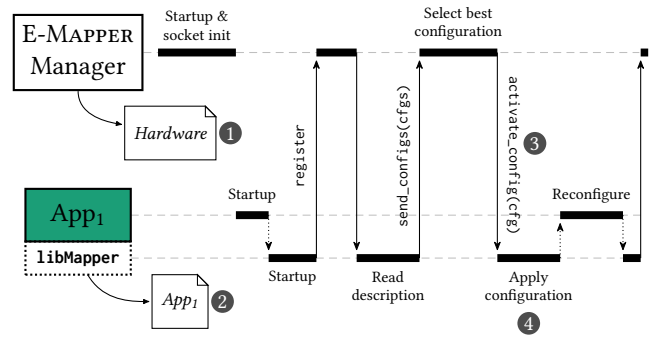
\begin{figure}[bt]
    \centering
    \begin{tikzpicture}
        \node[server] (server) {\mean{}\\Manager};

        \node[mapping, below right=.2cm and -.25cm of server] (hardware_map) {Hardware};
        \node[reference, right=1pt of hardware_map] {1};

        \draw[read mapping] (server) to[out=270, in=170] (hardware_map);

        \node[app, text opacity=1, fill=app1bg, below=1.3cm of server] (app1) {App\textsubscript{1}};
        \node[app, lib, align=center, below=0cm of app1] (lib1) {\libmean{}};
        \node[mapping, right=.655cm of lib1.south east, anchor=north] (map1) {App\textsubscript{1}};
        \node[reference, right=1pt of map1] {2};

        \node[dummy, right=5pt of server] (bserver) {};

        \node[dummy, right=1cm of bserver] (bstartup) {};
        \draw[active] (bserver |- server) -- (bstartup |- server) node[txt, midway, sloped, above] {Startup \&\\socket init};

        \node[dummy, right=.4cm of bstartup] (app1startup) {};
        \draw[active] (bstartup |- app1) -- (app1startup |- app1) node[txt, midway, sloped, above] {Startup};
        \draw[call] (app1startup |- app1) -- (app1startup |- lib1);

        \node[dummy, right=.6cm of app1startup] (lib1startup) {};
        \draw[active] (app1startup |- lib1) -- (lib1startup |- lib1) node[txt, midway, sloped, below] {Startup};
        \draw[communicate] (lib1startup |- lib1) -- (lib1startup |- server) node[txt, midway, sloped, above] {\texttt{register}};

        \node[dummy, right=.5cm of lib1startup] (register) {};
        \draw[active] (lib1startup |- server) -- (register |- server);
        \draw[communicate] (register |- server) -- (register |- lib1);

        \node[dummy, right=.8cm of register] (read_config) {};
        \draw[active] (register |- lib1) -- (read_config |- lib1) node[txt, midway, sloped, below] (read1) {Read\\description};
        \draw[read mapping] (lib1) to[out=300, in=180] (map1);
        \draw[communicate] (read_config |- lib1) -- (read_config |- server) node[txt, midway, sloped, above] (send1) {\texttt{send\_configs(cfgs)}};

        \node[dummy, right=1cm of read_config] (select_best) {};
        \draw[active] (read_config |- server) -- (select_best |- server) node[txt, midway, above, sloped] {Select best\\configuration};
        \draw[communicate] (select_best |- server) -- (select_best |- lib1) node[txt, midway, sloped, above] (send) {\texttt{activate\_config(cfg)}};
        \node[reference, above=0pt of send, xshift=2pt] {3};

        \node[dummy, right=.8cm of select_best] (reconfigure) {};
        \draw[active] (select_best |- lib1) -- (reconfigure |- lib1) node[txt, midway, below, sloped] (apply) {Apply\\configuration};
        \node[reference, below=-2pt of apply] {4};
        \draw[call] (reconfigure |- lib1) -- (reconfigure |- app1);

        \node[dummy, right=.8cm of reconfigure] (reconfigure_ack) {};
        \draw[active] (reconfigure |- app1) -- (reconfigure_ack |- app1) node[txt, midway, above, sloped] {Reconfigure};
        \draw[call] (reconfigure_ack |- app1) -- (reconfigure_ack |- lib1);

        \node[dummy, right=.2cm of reconfigure_ack] (done) {};
        \draw[active] (reconfigure_ack |- lib1) -- (done |- lib1);
        \draw[communicate] (done |- lib1) -- (done |- server);

        \draw[active] (done |- server) -- ++(.1cm, 0);

        \begin{pgfonlayer}{bg}
            \draw[gray, opacity=.5, densely dashed] (server) -- (done);
            \draw[gray, opacity=.5, densely dashed] (app1) -- (done |- app1);
            \draw[gray, opacity=.5, densely dashed] (lib1) -- (done |- lib1);
        \end{pgfonlayer}
\end{tikzpicture}
    \caption{
      Typical control flow between a managed application and the \mean{}
      resource manager.
    }
    \label{fig:control_flow}
\end{figure}

A typical control flow between a managed application and the \mean{} resource
manager is shown in \cref{fig:control_flow}. At application startup, the
\libmean{} library initializes, registers with the resource manager, and reads
the application description file, which contains an application identifier and
the list of operating points. Additionally, the library performs checks on the
application binary to identify potential configuration options. The collected
information is transmitted to the \mean{} manager, which thereon reevaluates the
current system load and resource allocations and selects new configurations for
all managed applications. These selected configurations are communicated back to
the individual application libraries, which adapt the applications accordingly.
Hence, the \libmean{} library responsible for the newly started application
(App\textsubscript{1} in the figure) also receives an \texttt{activate\_config}
message and performs the initial application configuration. Such messages can
occur anytime during an application's lifetime, potentially reconfiguring the
application multiple times over the course of its execution.

\subsection{\mean{} Application Types}
\label{sec:client_types}

\mean{} distinguishes between different types of applications to optimally
manage them within a heterogeneous processor. The main difference between the
types lies in the adaption options available to each type. Determining the
correct application type is done by the \libmean{} library during startup or
specified in the application's description file.

\subsubsection*{Static Applications}
\label{sec:client_types:static}

Applications without identified runtime adaptation mechanisms are labeled as
\emph{static}. Static applications are allocated to specific cores at runtime,
always according to an operating point selected by the manager. Depending on the
application description, the \libmean{} library might allocate specific
application threads to individual cores or restrict the entire application to a
subset of available processor cores. If applications run with more threads than
the cores allocated by \mean{}, the possible parallelism will be lower as
threads will have to time-multiplex the same processor core.

\subsubsection*{Scalable Applications}
\label{sec:client_types:scalable}

The \libmean{} library searches for supported runtime libraries for adaptive
execution, such as Intel TBB, OpenMP, and TensorFlow. If such libraries are
used, the application is marked as \emph{scalable}. \mean{} can dynamically
scale applications when resource allocations change. For example, if an
application is reallocated to two E-cores on an Intel Raptor Lake system, the
\libmean{} library not only adapts the CPU affinity but also instructs the
runtime library to use two threads, allowing applications to better adapt to new
configurations and reduce the interference due to resource overutilization.

\subsubsection*{Custom Applications}
\label{sec:client_types:custom}

Application developers can provide their own extensions to support more
application configuration options. For instance, we implemented a \libmean{}
library extension that enables application-specific scaling for Kahn Process
Networks (KPN)~\cite{khasanov18}. Instead of generically scaling the number of
threads, this extension scales specific \emph{parallel regions} inside the
application, leaving other parts unmodified. This allows fine-grained resource
allocations for different application phases.

Future work could investigate further extensions that react to the allocated
resources, such as using different algorithms or specialized code paths within
the application or handling ISA-extension differences between core types. These
configurations need to be application-specific and provided by the developer.
Our \libmean{} provides an interface to extend communication with the \mean{}
manager, easily supporting application-specific adaptations.

\subsection{\mean{} within the System Stack}
\label{sec:design:os_stack}

\mean{} aims to extend the current system stack rather than replace an existing
component. \mean{} relies on OS components such as the scheduler or performance
and power monitoring. \mean{}'s core responsibility requires a holistic view of
the system, similar to modern system management daemons such as systemd, OpenRC,
and launchd, we envision \mean{} to play a central role within user-space
management. Hardware description files (\cref{fig:system_overview:hardware_map})
would be provided by the system manufacturer or distribution while application
description files (\cref{fig:system_overview:app_map}) would be shipped together
with the applications. Managing these description files in a central file
hierarchy would allow a well-defined and user-extensible configuration database,
enabling easy adjustments and powerful resource management by \mean{}.

\subsubsection*{Making Applications \mean{}-ready}
\label{sec:design:ready}

To achieve the best results with \mean{}, applications should be equipped with a
description file tailored to the hardware they run on. For static or scalable
applications supported by \libmean{} out-of-the-box, no further adjustments are
needed. The library automatically identifies if the application can be
dynamically adapted and reconfigures it at runtime accordingly. Custom
application support, like for KPNs, requires changes to the application itself
for proper \mean{}-integration. If an application lacks a description file or
has an incomplete one, \mean{} can automatically learn the application's
behavior at runtime. The \emph{Online Configuration Refinement} algorithm is
described in detail in \cref{sec:runtime_dse}.

\section{\mean{} Resource Allocation}
\label{sec:selection_algorithm}

The core challenge addressed by \mean{} is the optimal allocation of resources
to applications. Thus \mean{}'s algorithm is designed to balance the resource
needs of applications with overall system energy efficiency.

\subsection{System Model}
\label{sec:selection_algorithm:system_model}

In \mean{}, a platform is equipped with a \emph{heterogeneous processor}
$\mathcal{P}$ with $m$ core types, represented as the vector
$\mathcal{P}[\vec\Theta] = [\Theta_1, \dots, \Theta_{m}]^T$. Each core type,
while sharing the same Instruction Set Architecture (ISA), exhibits distinct
per\-for\-man\-ce-energy characteristics and may have varying hardware
properties such as different ISA extensions or the number of supported hardware
threads. Since performance and energy characteristics of the multi-threaded
cores also vary depending on number of used hardware threads, our algorithm
ensures that applications are isolated at the core level, thus do not share
sibling hardware threads.

When an application $\sigma$ connects to \mean{}, it sends an application
description containing a list of \emph{operating points} $\sigma[\Phi]$. Each
operating point $\sigma[\phi_i] \in \sigma[\Phi]$ is defined by the required
resources $\vec{\theta}$, (normalized) utility $\upsilon$, and average power
consumption $\rho$, i.e.,
$\sigma[\phi] = \phi \langle \vec{\theta}, \upsilon, \rho \rangle$. Resource
requirements are specified at the core level; if an operating point uses
multiple hardware threads of a single processor core, this core is counted only
once in $\phi[\vec{\theta}]$. Operating points are assumed to be
Pareto-filtered, meaning each point is better than any other in at least one
parameter, such as fewer cores of a particular core type $\theta_k$, higher
utility $\upsilon$, or lower power consumption $\rho$.

For each operating point, \mean{} calculates the \emph{energy-utility cost}
$\phi[\zeta]$, an adaptation of the traditional Energy-Delay Product (EDP)
formula designed to balance energy efficiency with the performance impact of
applications~\cite{martin_ed2, general_edp}. Assuming that utility is inversely
proportional to delay, the energy-utility cost is defined as follows:
\begin{equation}
\phi[\zeta] = \left(\frac{\phi[\rho]}{\phi[\upsilon]}\right) \cdot \left(\frac{1}{\phi[\upsilon]}\right)
\end{equation}

\subsection{Optimization Problem}

The \mean{} resource allocation algorithm selects one operating point per
application in a way that minimizes the overall system energy-utility cost. This
optimization problem can be formulated as:
\begin{subequations}
\begin{align}
    \text{\textbf{minimize} } & \sum_{\sigma \in \Sigma} \sigma[\phi^*][\zeta], \label{eq:design:energy}\\
    \text{\textbf{subject to} } & \sum_{\sigma \in \Sigma} \sigma[\phi^*][\vec{\theta}] \leq \mathcal{P}[\vec{\Theta}]. \label{eq:design:resource}
\end{align}
\end{subequations}
\cref{eq:design:energy} minimizes the sum of the energy-utility costs of each
application's selected operating points. The constraint in
\cref{eq:design:resource} ensures that the total resource demand from all
selected operating points does not exceed the available resources for each
processor core type.

This optimization problem is equivalent to the \emph{multiple choice
multi-dimensional knapsack problem} (MMKP)~\cite{knapsack_book}. In MMKP, all
items have a \emph{scalar} value and a \emph{multi-dimensional} weight, and they
are divided into several groups. The goal is to select a single item from each
group (\emph{multiple-choice}) so that the overall value is maximized and the
overall weight does not exceed the maximum allowed weight at each dimension.

In our optimization problem, each application's operating points represent items
in a group. The goal is to select one operating point per application (one item
per group) such that the overall value is minimized. Here, the \emph{weight} is
the number of used processors of each type, and the \emph{value} of the item is
represented as a negative energy-utility cost.

Given that MMKP is NP-hard~\cite{puchinger10}, finding an optimal solution
within a reasonable time, especially for a large number of applications and
operating points, is computationally challenging. Since our resource allocation
algorithm is used at runtime, \mean{} employs a state-of-the-art approximate
algorithm based on Lagrangian relaxation, which solves the optimization problem
with relaxed constraints and then selects the resultant operating points that
fulfill the resource constraints. The detailed description of the algorithm is
provided in~\cite{wildermann14,wildermann15}.

\subsubsection*{Limitations}

It is possible that the resource allocation algorithm does not find any suitable
operating point for some of the applications due to prior allocations. This
situation may occur if the number of managed applications exceeds the number of
available resources. In such cases, \mean{} temporarily relaxes the constraint
in \cref{eq:design:resource}, allowing applications to execute in co-allocation.
Since co-alloca\-tion adversely affects the performance of the co-allocated
applications, \mean{} does not perform performance monitoring, as discussed in
the next section.

\section{Explorating Operating Points at Runtime}
\label{sec:runtime_dse}

\begin{figure*}[htbp]
  \centering
  \includegraphics[width=1\textwidth]{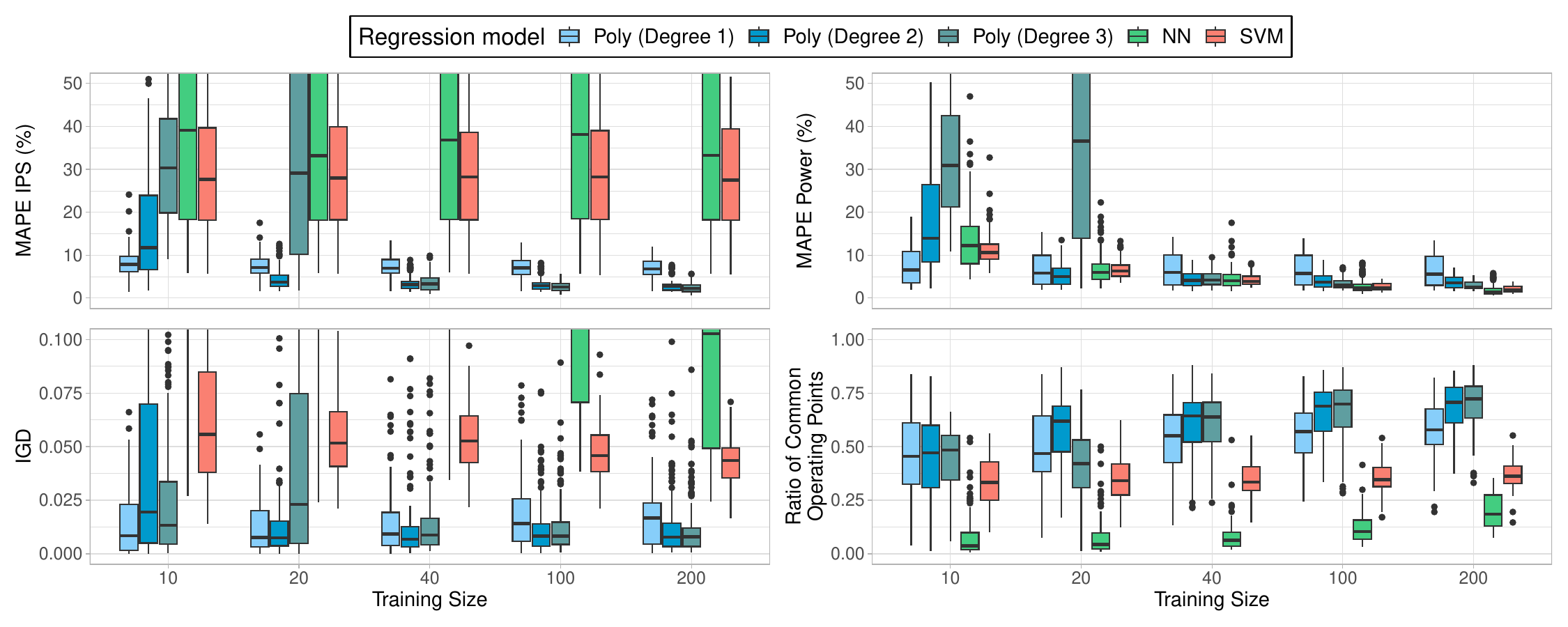}
  \vspace{-6mm}
  \caption{
    Evaluation of several regression models in terms of Mean Absolute Percentage
    Error (MAPE) for IPS and Power (lower is better), Inverted Generational
    Distance (IGD) (lower is better), and the ratio of common operating points
    in the Pareto-front (higher is better) across 15 different applications on
    the Intel Raptor Lake Core i9-13900K.
  }
  \vspace{-3mm}
  \label{fig:runtime_dse:regression}
\end{figure*}

Traditionally, operating points for HAM approaches have been generated at design
time, leveraging substantial computational resources to identify Pareto-optimal
configurations~\cite{pourmohseni20}. However, in practical PC scenarios, the
operating system may not have predefined operating points, or the available ones
might be imprecise due to variations in hardware, despite using similar
architectures. Runtime exploration of operating points represents a synergy
between Design Space Exploration (DSE) and Runtime Resource Management. This
integration poses several challenges, primarily the need to identify effective
operating points quickly. Once identified, the algorithm continues refining the
Pareto front as new measurements become available.

During exploration, online measurements determine the utility and power
consumption of the running applications. If applications do not provide their
own utility metrics, we use Instructions Per Second (IPS) as a generic measure
of utility. Performance metrics, such as IPS and power, are not constants; they
fluctuate due to measurement noise and varying application stages. This
variability requires periodic re-evaluation to ensure the robustness and
reliability of solutions. Furthermore, the search for near-optimal
configurations must be seamlessly integrated within the runtime resource
management of the \mean{} manager. This integration ensures that while exploring
new operating points, the selected configurations do not compete with other
concurrently running applications on processor cores. It also requires that the
resource management algorithm allocates sufficient resources to new
applications, allowing them substantial solution space for exploration without
significantly undermining the performance of existing applications.

To address these challenges, we have enhanced the resource allocation algorithm
in \mean{} to include the exploration of operating points directly during
application execution. This process involves continuous performance monitoring
and employing a regression model to dynamically adjust operating points,
ensuring robust, Pareto-optimal configurations. Our algorithm effectively
balances new operating point exploration with efficient resource allocation,
optimizing overall system performance.

\subsection{Runtime Performance and Power Monitoring}
\label{sec:runtime_dse:monitoring}

Accurate monitoring of runtime performance metrics is essential for a good
approximation of Pareto fronts. To reliably measure performance values such as
IPS, we employ the Linux performance monitoring subsystem
\texttt{perf}\footnote{Most of its functionality is provided by the system call
\texttt{perf\_event\_open}.}, which allows monitoring various
performance-related hardware and software metrics, such as retired instructions,
cache misses, etc. With the proper configuration, \texttt{perf} automatically
multiplexes measurements for different applications.

For assessing power consumption, we use built-in power sensors available in
modern systems, such as the \texttt{RAPL} (Running Average Power Limit) counters
on Intel machines. However, these sensors generally measure the total energy
consumption of the system rather than per application. To address this
challenge, we build atop EnergAt~\cite{energat}, which monitors \texttt{RAPL}
counters alongside the execution metrics of each thread across the hardware
threads.

Since EnergAt does not support different core types in heterogeneous CPUs, we
introduce power coefficients between core types ($P^{P} = \gamma \cdot P^{E}$,
determined statically) to attribute total energy consumption
($E_{\Delta}^{CPU}$) to P-cores ($E_{\Delta}^{P}$) and E-cores
($E_{\Delta}^{E}$):
\begin{equation}
E_{\Delta}^{CPU} = E_{\Delta}^{P} + E_{\Delta}^{E}  = T_{total}^{P} \cdot P^{P} + T_{total}^{E} \cdot P^{E} 
\end{equation}
where $T_{total}^{P|E}$ is the sum of CPU time for all threads running on
P-cores and E-cores, respectively. After approximating $E_{\Delta}^{P}$ and
$E_{\Delta}^{E}$ values, we can employ the EnergAt methodology to further
attribute energy to the applications running on homogeneous subsets of cores.

Considering the inherent variability in measured IPS and power, we implement an
exponential moving average (EMA) to stabilize these metrics, accommodating
recognition of changes in application stages and their corresponding power and
performance characteristics. The EMA is updated as follows:
\begin{equation}
    value_{new} = value_{measured} \cdot \alpha + value_{old} \cdot (1 - \alpha)
\end{equation}
where $\alpha$ (\mean{} uses $0.1$) is a smoothing factor. This formula smooths
short-term fluctuations while adapting to significant shifts in application
behavior, ensuring accurate and responsive performance and power profiling.

\subsection{Selection of the Regression Model}
\label{sec:runtime_dse:regression}

We evaluated several regression models for predicting IPS and power consumption
for unexplored operating points. Each operating point's configuration is
characterized with a vector that includes the number of cores of each type and
their hardware thread usage. For instance, on the Intel Raptor Lake platform,
the configuration vector includes the number of E-cores, P-cores using one
hardware thread, and P-cores using both hardware threads. The models assessed
were Polynomial Regression (degrees 1 to 3), Neural Networks (NN), and Support
Vector Machines (SVM). We used pre-measured configurations across 15
applications on the Intel Raptor Lake Core i9-13900K as the data sets. Each
model was evaluated using training subsets of different sizes across 10 randomly
generated seeds for robustness.

The first two plots in Figure~\ref{fig:runtime_dse:regression} show the Mean
Absolute Percentage Error (MAPE) for predicted IPS and power values. Polynomial
regression models improved in accuracy for both IPS and power as the training
size increased. Higher-degree polynomial models achieved greater accuracy at
larger training sizes, albeit requiring more data points to converge.
Conversely, NN and SVM models performed better in predicting power values at
larger training sizes but were significantly worse in predicting IPS compared to
polynomial models.

Furthermore, we compare the predicted Pareto fronts with a reference Pareto
front (derived from the measured configurations) using Inverted Generational
Distance (IGD)~\cite{coello04} and the ratio of common operating points. IGD
measures the average distance from points on the reference Pareto front to the
nearest point on the generated front, assessing the coverage of the generated
front relative to the reference one. The bottom two plots in
Figure~\ref{fig:runtime_dse:regression} show that polynomial models consistently
outperformed SVM and NN in aligning with the reference Pareto front.
Particularly, polynomial regression models of degrees 2 and 3 outperformed the
first-degree model in approximating the reference front. Both second and
third-degree models produced similar accuracy, but the second-degree model was
more efficient, requiring only 20 training points to generate robust Pareto
fronts. Based on this efficiency, we selected the second-degree polynomial
regression model for our runtime exploration approach.

\subsection{Runtime Exploration Algorithm Design}
\label{sec:runtime_dse:exploration}

As discussed earlier, it is critical to integrate the runtime exploration of
operating points seamlessly into the resource allocation algorithm. The chosen
points for measurement should not overlap with processor cores used by
concurrently running applications. Simultaneously, the resource allocation
algorithm should allocate sufficient resources to ensure effective exploration
without adversely affecting the performance of other applications.

We categorize the maturity of an application's operating points into three
stages: (1) \textit{Initial} stage, where there are insufficient measured
operating points making approximations unreliable; (2) \textit{Refinement}
stage, characterized by an intermediate number of measured points but still
limited accuracy in approximations; and (3) \textit{Stable} stage, at which
sufficient operating points have been explored, enabling reliable
approximations.

Upon invocation, the resource allocation algorithm, as detailed in
Section~\ref{sec:selection_algorithm}, generates the current Pareto front using
both measured and approximated operating points. This allocation determines the
set of cores assigned to each application. If there are unassigned cores, the
exploration algorithm allocates these to applications in the Initial and
Refinement stages, allowing them to explore a broader configuration space.
Applications in the Stable stage execute on the designated cores provided by the
resource allocation algorithm without further configuration adjustments. For
applications in the Initial and Refinement stages, specific exploration
techniques are employed within the set of remaining cores:

In the \emph{Initial} stage, the selection heuristic for the next operating
point is based on the configuration vector, choosing the configuration furthest
from the measured configurations to maximize exploration diversity. In the
\emph{Refinement} stage, the heuristic is based on approximated IPS and power
values and uses an auxiliary regression, which includes a "zero" configuration
(zero power and zero IPS) to anchor the model. The heuristic selects
configurations based on the largest discrepancies between the main and the
auxiliary regression models. These discrepancies are calculated as the geometric
mean of the relative differences in IPS and power values. If both models predict
negative IPS or power values for some configurations, discrepancies are
calculated relative to a zero value to increase the likelihood of measuring
configurations that exhibit such anomaly predictions. After a fixed number of
measurements, the exploration process repeats until the application progresses
to the \emph{Stable} stage. When applications are in the Stable stage, the
resource allocation algorithm is invoked again after a larger number of
measurements to reassess the current allocation and potentially switch to
another one.

\section{Evaluation}
\label{sec:evaluation}

To demonstrate that \mean{} enhances the management of energy-aware tasks on
heterogeneous processors, we conducted an extensive evaluation on two different
systems. We explicitly choose systems with distinct characteristics to underline
that our approach is generic and works across various types of heterogeneous
hardware.

\subsection{Evaluation Setup}
\label{sec:evaluation:setup}

We evaluate our approach on an Arm-based Odroid XU3-E board~\cite{xue}. The
board features a Samsung Exynos 5422 processor, which implements an Arm
big.LITTLE architecture with two core islands, a four-core A15 (big) island and
a four-core A7 (LITTLE) island. The system is equipped with \qty{2}{\giga\byte}
of memory and energy sensors for the core islands, the memory, and the
integrated graphics card. We run a custom-compiled Linux~6.6 kernel on the
board, with full support for the Linux Energy-Aware-Scheduler (EAS), a Linux'
built-in optimized application placement strategy~\cite{linux_eas,
linux_eas_docs}.

The second system is an Intel Raptor Lake Core i9-13900K, one of Intel's latest
heterogeneous processors. It consists of 8 high-performance P-cores, each
supporting SMT, and 16 energy-efficient E-cores that do not support SMT. The
system is equipped with \qty{128}{\giga\byte} of memory. For energy
measurements, we use the integrated RAPL counters, which have been proven
accurate in fine-grained energy measurements~\cite{smejkal_eteam,
Hahnel.etal_12_MeasuringEnergyConsumption,
Colmant.etal_15_ProcesslevelPowerEstimation,
Schone.etal_19_EnergyEfficiencyFeatures,
Hackenberg.etal_15_EnergyEfficiencyFeature}.

We run a custom-compiled Linux~6.4 kernel, based on the default Debian Testing
kernel version, extended with a patch-set adding preliminary support for Intel
Thread Director (ITD)~\cite{linux_classes}. We further extended the patch-set to
make the ITD classification of threads and the reference IPC per class
determined by the hardware available to user-space. Inspired by Saez et
al.~\cite{saez2022evaluation} we implemented a version of \mean{} that uses ITD
classifications to allocate processor cores to application threads.

For both platforms we use the \texttt{performance} frequency governor and limit
the maximum frequencies to prevent thermal throttling, allowing the cores to run
at maximum frequencies for the entire benchmark execution. On the Raptor Lake
system, we selected \qty{4.6}{\giga\hertz} for the P-cores and
\qty{3.8}{\giga\hertz} for the E-cores, while on the Odroid system, we chose
\qty{1.2}{\giga\hertz} for the LITTLE and \qty{1.8}{\giga\hertz} for the big
cores.

\subsection{Benchmarks}
\label{sec:evaluation:applications}

\begin{figure*}[htbp]
  \centering
  \includegraphics[width=1\textwidth]{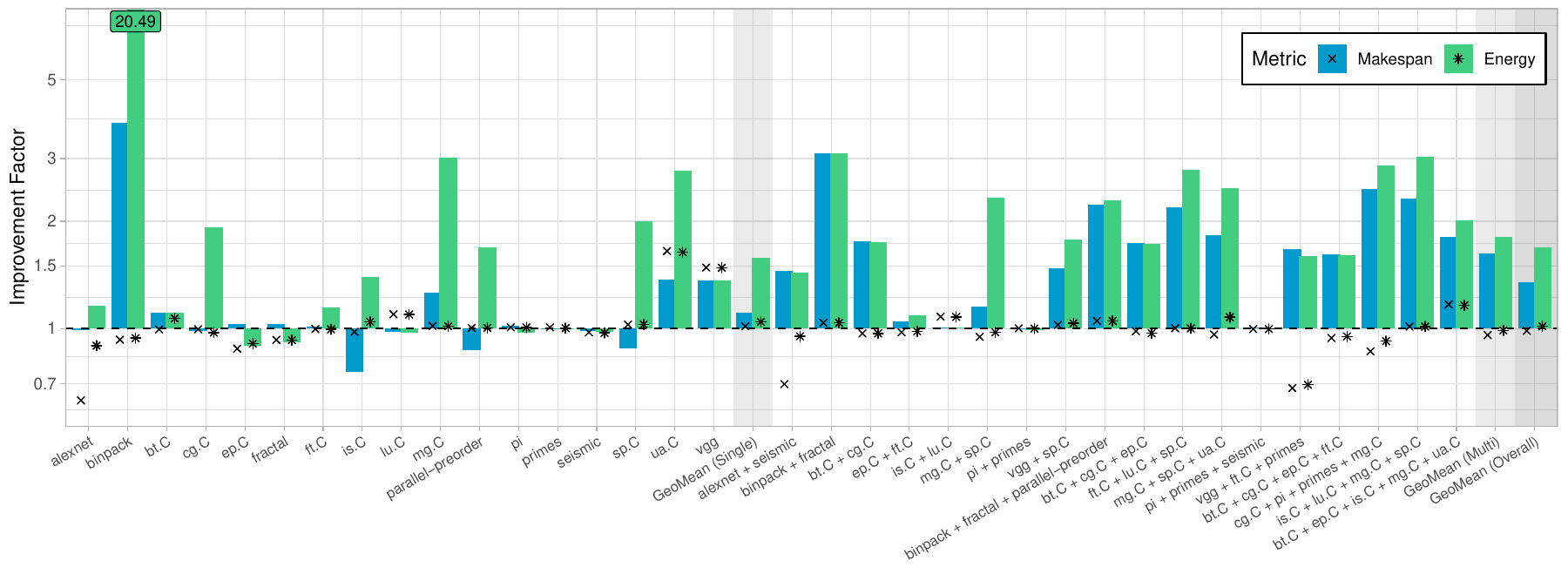}
  \vspace{-6mm}
  \caption{
    Relative improvement factor \mean{} and Intel Thread Director over CFS on
    the Intel Raptor Lake Core i9-13900K (higher y-value is better). \mean{}
    results are shown as bars, Intel Thread Director results are shown as
    points.
  }
  \vspace{-3mm}
  \label{fig:evaluation:raptor}
\end{figure*}

We use different sets of applications to evaluate the features of \mean{}. To
test dynamic adaptability, we use the OpenMP implementations of the NAS Parallel
Benchmarks~\cite{npb}, version 3.4.2. Since the Intel platform is more powerful
than the Odroid board, we use different classes of the benchmarks for each
system: class \emph{A} for Odroid and class \emph{C} for Intel.

On the Intel Raptor Lake platform, we also evaluate a selection of Intel Thread
Building Blocks (Intel TBB)~\cite{pheatt2008inteltbb} benchmarks and two
TensorFlow~\cite{developers2022tensorflow} applications. We chose the benchmarks
\texttt{binpack}, \texttt{fractal}, \texttt{parallel-preorder}, \texttt{pi},
\texttt{primes}, and \texttt{seismic} from the official Intel TBB repository as
they cover a wide spectrum of the building blocks of the framework. TensorFlow,
an open-source framework for machine-learning algorithms, is also included in
our evaluation. We implemented a \mean{}-enabled version of TensorFlow
Lite~\cite{tflite} that can dynamically scale its parallelism at runtime. We
evaluate two models, VGG~\cite{simonyan2014very} and
AlexNet~\cite{krizhevsky2017imagenet}, used for image recognition.

Additionally, we use two embedded KPN applications to test the custom extensions
of \mean{}: \texttt{mandelbrot} for calculating the Mandelbrot
set~\cite{khasanov18}, and \texttt{lms} implementing Leighton-Micali
Signatures~\cite{lms}. Both applications are used in two versions: one with a
static application topology (annotated with \texttt{static}) and another with
implicit data-parallelism in KPNs~\cite{khasanov18}, used to demonstrate dynamic
adaptation. Since KPN applications are targeted to embedded platforms, we chose
to evaluate them only on the Odroid system.

\subsection{Intel Raptor Lake}
\label{sec:evaluation:raptor}

For our first experiment, we measure the execution of  our selected benchmark
collection on the Raptor Lake system. We evaluate both single benchmarks and
parallel executions of multiple benchmarks. Every scenario is executed with
resource management by \mean{} and ITD-based resource allocation as described
earlier. With \mean{}, we use application description files generated from a
design space exploration prior to the actual measurement. The ITD-based
allocation does not require this step, instead uses hardware-provided
classification information for resource allocation. We also measure scenarios
using the Linux built-in work-distribution techniques as a baseline.

For each scenario, we record the overall execution time (makespan) and total
energy consumption. Each scenario is measured ten times, and the average results
are reported. \Cref{fig:evaluation:raptor} shows the results for all benchmarks,
including geometric means for single application scenarios, multiple application
scenarios, and all scenarios combined. The results are presented as improvement
factors over the baseline; an improvement factor above 1 indicates faster
execution or lower energy consumption, while below 1 indicates the opposite.

\subsubsection*{Single Application}
\label{sec:evaluation:raptor:single}

For single application scenarios, benchmarks generally benefit from being
managed by \mean{}. The geometric mean improvement factor for \mean{} is
\qty{1.11}{} for execution time and \qty{1.57}{} for energy consumption. There
are two notable cases to highlight. First, the \texttt{binpack} benchmark from
Intel TBB shows a \qty{10}{\times} higher IPS with low thread counts. \mean{}
leverages this behavior by scaling the application accordingly, while the
baseline and ITD-based manager cannot, thus resulting in extreme improvement.
Second, benchmarks \texttt{is}, \texttt{sp}, and \texttt{parallel-preorder} run
longer with \mean{} but consume less energy due to \mean{}'s optimization
algorithm, which balances power consumption and utility.

The ITD-based resource allocation shows only minor differences from the
baseline, with overall improvement factors of \qty{1.01}{} for execution time
and \qty{1.04}{} for energy consumption. ITD improves the \texttt{ua} and
TensorFlow \texttt{vgg} benchmarks by allocating P-cores to more active threads.
However, it misclassifies threads for \texttt{alexnet}, leading to suboptimal
performance.

\subsubsection*{Multiple Applications}
\label{sec:evaluation:raptor:multi}

For scenarios with multiple applications running in parallel, \mean{} provides
significant benefits. The average improvement factor is \qty{1.62}{} for
execution time and \qty{1.81}{} for energy consumption. Most applications show
improvements, with few exceptions matching the baseline.

Managing resources based on ITD information did not yield the expected
improvements. Most scenarios only achie\-ved similar performance to the
baseline, resulting in an overall improvement factor of \qty{0.95}{} for
execution time and \qty{0.98}{} for energy consumption when multiple
applications ran in parallel. Some minor improvements were observed in scenarios
like \texttt{is} and \texttt{lu} running in parallel, and when five applications
run simultaneously. However, combinations like \texttt{vgg} and \texttt{ft} and
\texttt{primes} showed worse results, highlighting classification issues.

\subsection{Odroid XU3-E}
\label{sec:evaluation:odroid}

\begin{figure*}[htbp]
  \centering
  \includegraphics[width=1\textwidth]{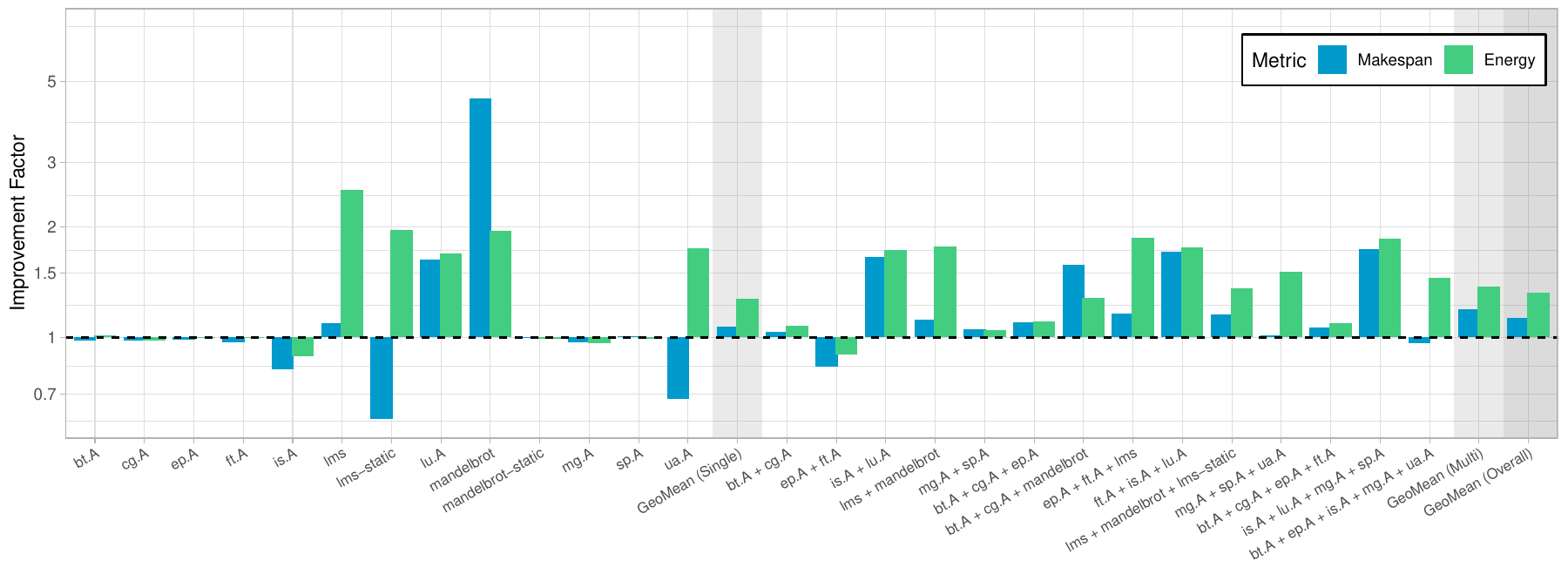}
  \vspace{-6mm}
  \caption{
    Relative improvement factor \mean{} over EAS on the Odroid XU3-E (higher
    y-value is better).
  }
  \vspace{-3mm}
  \label{fig:evaluation:odroid}
\end{figure*}

To demonstrate \mean{}'s effectiveness on different hardware, we conducted a
second experiment on the Odroid XU3-E platform, measuring single and multiple
application scenarios. The baseline for this platform uses the Linux
Energy-Aware Scheduler (EAS), which leverages the power model of the
heterogeneous processor. \Cref{fig:evaluation:odroid} shows the results of this
experiment.

\subsubsection*{Single Application}
\label{sec:evaluation:odroid:single}

The results for single OpenMP benchmarks are similar to the Intel Raptor Lake
experiment. The applications either perform similarly to the baseline or show
improved energy consumption with longer execution times (e.g., \texttt{ua}).
The \texttt{is} benchmark shows a significant increase in
execution time as well as an energy consumption increase. \mean{} picks a
sub-optimal configuration for this application leading to this negative
effects. In contrast, \texttt{lu}, a long-running benchmark, benefits
significantly from \mean{} in both energy consumption and execution time.

KPN applications show significant improvements with \mean{}. \mean{} can
successfully use the application knobs to tune the application to the available
hardware resources, resulting in lower energy consumption for \texttt{lms} and
improvements in both metrics for \texttt{mandelbrot}. The static version of
\texttt{mandelbrot} behaves the same as the baseline. The static version of
\texttt{lms} behaves similar to \texttt{ua} and shows a lower energy
consumption at the cost of a longer makespan. On average, \mean{} improved
execution time by a factor of \qty{1.07}{} and energy consumption by
\qty{1.27}{} for single application scenarios.

\subsubsection*{Multiple Applications}
\label{sec:evaluation:odroid:multi}

When running multiple applications in parallel, \mean{} significantly enhances
overall system performance. The geometric mean improvement factor is
\qty{1.20}{} for execution time and \qty{1.38}{} for energy consumption. Most
scenarios show improvements, with only the combination of \texttt{ep} and
\texttt{ft} suffering in execution time and energy consumption compared to the
baseline.

\subsection{Runtime Exploration of Operating Points}
\label{sec:runtime_refinement}

\begin{figure*}[htbp]
  \centering
  \includegraphics[width=1\textwidth]{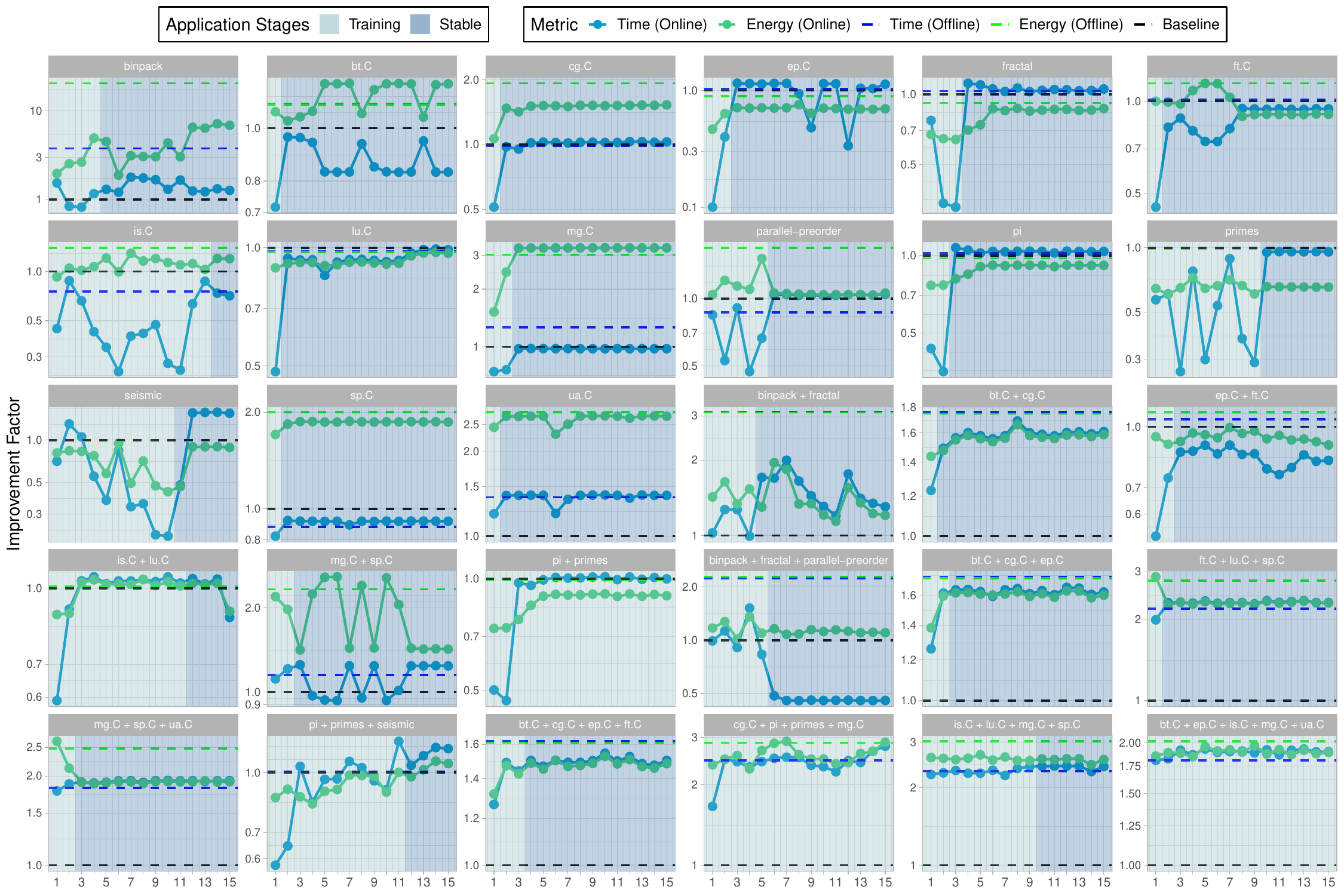}
  \vspace{-6mm}
  \caption{
    Runtime exploration of operating points across 30 different scenarios using
    the Intel Raptor Lake Core i9-13900K.
  }
  \vspace{-3mm}
  \label{fig:evaluation:learn}
\end{figure*}

This section demonstrates \mean{}'s behavior with runtime exploration of
operating points on the Intel Raptor Lake Core i9-13900K across 15
single-application and 15 multi-application scenarios.

Each scenario runs 15 times, starting with no predefined operating points.
Subsequent runs use data from prior runs to continue exploration. Applications
remain in the initial stage until 6 operating points are measured, move to the
refinement stage until 25 operating points with at least 20 measurements each
are gathered. The measurement period is \qty{50}{ms}, discarding the first three
measurements after reconfiguration to allow applications to adjust.
Figure~\ref{fig:evaluation:learn} shows the results. The "Training" stage
signifies ongoing exploration in at least one application, while the "Stable"
stage indicates all applications start from a stable stage.

Not all scenarios reach the stable stage within 15 executions. While all
scenarios with single applications reach the stable stage, the three
multi-application scenarios do not. These three scenarios involve short-running
applications alongside longer-running ones. We suspect the resource distribution
between the exploring and the stable stage applications is the issue;
applications in the exploration stages do not get enough resources to progress
to the stable stage. We believe that further refinement of the runtime
exploration algorithm could reduce the likelihood of such situations.

During the training stages, performance and energy consumption fluctuate, and
upon reaching the stable stage, the results stabilize, though \mean{} continues
to measure and refine operating points. Results are usually worse than those
with offline-generated operating points, but for most single-application
scenarios, online training gets close to the original results, indicating that
online exploration effectively identifies optimal or near-optimal
configurations. In multi-application scenarios, results are generally better
than the CFS baseline. A notable exception is the \texttt{binpack + fractal +
parallel-preorder} scenario, where runtime measurement and prediction result in
poor execution time due to a bad estimation for the \texttt{binpack} benchmark.
These negative results suggest that the runtime exploration heuristics could be
further improved to find better configuration points.

\section{Related Work}
\label{sec:related_work}

Optimizing the mapping of applications onto heterogeneous multi-core systems is
a well-known problem in embedded systems~\cite{singh13survey}. The mapping
approaches can be classified by the decision time into static mapping, runtime
mapping, and hybrid application mapping techniques.

\subsubsection*{Static Mapping Approaches}

Static mapping approaches are among the most widely studied approaches. In
static mapping, resources are allocated prior to application execution,
typically during compile or design time. This approach leverages static code
analysis, often supplemented by execution traces, profiling data, and models of
the target hardware, to determine the optimal resource allocation.

Various strategies are employed to find optimal static mappings. These include
meta-heuristic techniques like evolutionary algorithms for design-space
exploration (DSE)~\cite{alexandrescu2011genetic,quan2014biasedelitist,
kang2012multi}, as well as formulations of well-known problems, like ILP or SMT,
which are solved using dedicated solvers~\cite{malik2018satisfiability}.
Additionally, simpler heuristics based on domain-specific knowledge are also
used~\cite{castrillon2012,singh2013accelerating,brunet2013buffer,khasanov21}.

Since static mappings are specified at design time, they allow for more
extensive calculations to find an optimal allocation than any runtime approach.
This enables complex analysis methods and can produce binaries that are easy to
deploy since they do not necessarily need any runtime support. However, static
mapping approaches cannot adapt to dynamic scenarios, such as varying system
loads or the presence of other concurrently executing applications.

\subsubsection*{Dynamic Mapping Approaches}

On the other end of the spectrum lie dynamic mapping approaches. In these
methods, the decision regarding resource allocation is deferred until runtime,
where a simple  heuristic typically determines the assignment. In most cases,
the mapping decision is also combined with task scheduling within single
processing elements (PEs) and embedded within the operating
system~\cite{brandenburg2008scalability,zhuravlev2012survey}. As mentioned in
the \Cref{sec:introduction}, modern OS schedulers are evolving to better manage
heterogeneous multi-core processors, aiming to improve energy efficiency. These
schedulers employ techniques such as using CPU energy models~\cite{linux_eas} or
leveraging hardware-assisted tools such as Intel's Thread
Director~\cite{linux_classes,windows11_scheduler,linux_td,saez2022evaluation,
bilbao2023flexible}. However, dynamic approaches are inherently limited in their
optimization space. While they are more adaptive, the applied heuristics must
remain lightweight to ensure efficient execution. This requirement often
restricts them to operating with limited information.

\subsubsection*{Hybrid Mapping Approaches}

To address the limitations of static and dynamic mapping methods, a trend for
\emph{hybrid} application mapping (HAM) approaches has
emerged~\cite{pourmohseni20}. HAM strategies leverage extensive analyses from
static methods at compile time while deferring final decision-making to runtime.
This class of approaches ensures efficiency while retaining flexibility to adapt
to dynamic system loads. During the design stage, hybrid approaches employ
sophisticated DSE methodologies to characterize the impact of different resource
allocations on the non-functional properties of the mapping. The outcome of DSE
is a set of (possibly incomplete) mapping options, referred to as
\emph{operating points}, each characterized by the required amount and type of
resources and the obtainable execution properties. Approaches such
as~\cite{massari14,singh2013accelerating,onnebrink19} use heuristics for
efficient exploration of Pareto-optimal configurations, while
others~\cite{ascia04,mariani12,daarm,quan2015hybrid} apply evolutionary
algorithms. Other approaches also distinguish different application
scenarios~\cite{quan2015hybrid,schranzhofer2010dynamic,schor2012scenario,
spieck22}.

At runtime, the resource managers utilize the generated operating points and
partition resources among multiple, concurrently executed applications.
Algorithms may map applications on the platform one at a
time~\cite{singh2013accelerating,daarm}, or in a joint manner by formulating the
problem via a Multiple-choice Multidimensional Knapsack Problem
(MMKP)~\cite{knapsack_book}, and using ILP solvers~\cite{bini11}, or fast
knapsack heuristics based on Pareto algebra principles~\cite{shojaei13},
Lagrangian relaxation~\cite{wildermann14,wildermann15,spieck22}, or greedy
heuristics~\cite{ykman06}. Similar approaches are also discussed for cluster
scheduling in cloud environments, where the dynamic distribution of jobs in a
heterogeneous cloud environment is solved using ILP
solvers~\cite{tumanov_tetrisched}.

The aforementioned approaches generate \emph{spatial} multi-application
mappings, i.e., they only generate a mapping for the current running set of
applications and do not consider the changes in the workload nor further
optimize the execution. To address it, some works~\cite{khasanov_date20,
khasanov21,khasanov24} proposed generating \emph{spatio-temporal} mappings at
runtime instead. However, it is important to note that while these approaches
offer significant improvements in terms of energy efficiency and meeting
real-time constraints, they often postpone the execution of the application to
select the more efficient operating later. Such a strategy, while suitable for
real-time systems, is less ideal for regular desktop computers where users
expect immediate application responsiveness.

\subsubsection*{Application Adaptivity}

A common limitation in many of the mapping approaches discussed is their focus
primarily on thread-to-core pinning, without fully exploiting the adaptability
potential of applications. Beyond mere assignment of threads to processor cores,
certain applications possess additional \emph{knobs} that allow them to adjust
their internal configurations dynamically at runtime. For instance, some
applications have the capability to modify their topology, alter the
parallelization degree of data-parallel regions, or switch between different
internal algorithms based on runtime conditions~\cite{khasanov18,
schor14_adapnet}. Another scaling approach involves distributing work within
OpenMP applications in a heterogeneous-hardware aware
fashion~\cite{saez202openmp}. Despite these possibilities, modern operating
system often fail to leverage these adaptive features, missing opportunities to
optimize application performance and resource utilization.

\section{Conclusion}
\label{sec:conclusion}

This paper introduces \mean{}, a resource management system designed for
energy-aware applications on heterogeneous processors. \mean{}'s key innovations
include a uniform interface to a global resource manager for passing high-level
application descriptions and an allocation algorithm that balances performance
and energy consumption. \mean{} supports a broad spectrum of application models,
from static to scalable types, and even custom-specific ones with unique
adaptivity features like reconfiguration options and algorithmic variations.
Crucially, \mean{} does not require detailed knowledge of each application's
adaptivity aspects. Instead, it receives only the essential information needed
for efficient resource allocation (resource requirements and non-functional
characteristics), with the rest handled on the client side. This flexible design
allows for various management scenarios.

Evaluations on two different heterogeneous systems show\-ed that \mean{} can
significantly improve application execution, both when given exclusive access to
resources and when competing with concurrent applications. We reported
improvements in terms of execution time and energy consumption of \qty{25}{\%}
and \qty{40}{\%} for the Intel Raptor Lake Core i9-13900K and \qty{12}{\%} and
\qty{25}{\%} for the Odroid XU3-E. For embedded KPN applications with custom
adaptivity knobs, we demonstrated \mean{}'s extensibility in providing
fine-grained resource adaptations. Finally, we showed that even in the absence
of application descriptions, \mean{} can efficiently manage unknown applications
through its runtime exploration approach, which predicts application behavior
based on few samples.

In summary, this work highlights the value of a simple yet effective interface
to express application characteristics to the OS. By communicating operating
points annotated with resource requirements and non-functional characteristics,
\mean{} considerably improves the system's overall energy-awareness. We advocate
for the development of a uniform application-resource-manager interface to
enhance energy efficiency across diverse applications and systems.

\section*{Acknowledgement}
\label{sec:acknowlegdement}

We would like to thank Andr\'{e}s Goens and Marcus H\"{a}hnel for early
discussions, as well as Dylan Gageot, Fabius Mayer-Uhma, and Marc Dietrich for
their contribution in supporting Kahn Process Networks and TensorFlow
applications. In addition, we thank the anonymous reviewers from ASPLOS 24 for
their valuable input and feedback, which led to the introduction of the runtime
exploration component. The authors acknowledge the financial support by the
Federal Ministry of Education and Research of Germany in the programme of
``Souver\"an. Digital. Vernetzt.'' (joint project 6G-life, project number
16KISK001K) and the E4C project (16ME0426K). This work also received funding
from the EU Horizon Europe Programme under grant agreement No 101135183
(MYRTUS). Views and opinions expressed are however those of the author(s) only
and do not necessarily reflect those of the European Union. Neither the European
Union nor the granting authority can be held responsible for them.

\vfill{}
\pagebreak{}

\bibliographystyle{ACM-Reference-Format}
\bibliography{paper}

\end{document}